\documentclass[aps,twocolumn,showpacs,preprintnumbers]{revtex4}

\usepackage{graphicx}  
\usepackage{subfigure}
\usepackage{multirow}
\usepackage{amsmath}

\usepackage{fancyhdr}
\usepackage{longtable}
\usepackage{parskip}
\usepackage[T1]{fontenc}
\usepackage{dcolumn}   

\usepackage{bm}        
\usepackage{amsfonts}  
\usepackage{amsmath}   
\usepackage{amssymb}   

\usepackage{natbib}

\newcommand{\pwisein}{\left\{ \begin{array}{ll}}
\newcommand{\pwiseout}{\end{array}\right.}

\begin{document}

\title{New Challenges in Plasma Accelerators: Final Focusing for Wakefield Colliders}

\author{Keegan Downham}

\affiliation {\it University of California, Santa Barbara, CA 93106}
\affiliation{\it SLAC National Accelerator Laboratory, Menlo Park, CA 94025}

\author{Spencer Gessner}

\affiliation{\it SLAC National Accelerator Laboratory, Menlo Park, CA 94025}

\author{Lewis Kennedy}

\affiliation{\it CERN, Geneva, CH}

\author{Rogelio Tom\'{a}s}

\affiliation{\it CERN, Geneva, CH}

\author{Andrei Seryi}
\affiliation{\it Old Dominion University, Norfolk, VA 23529}

\date{January 31, 2026}

\begin{abstract}  

The focusing of particle beams for collider experiments is crucial 
for maximizing the luminosity and thus the discovery potential of these 
machines. In recent years, plasma wakefield acceleration has emerged as a 
leading candidate for achieving higher energy collisions with smaller facility footprints due to the large accelerating gradients in the plasma. This higher beam energy poses significant
challenges for the final focusing system of the collider. Here, we discuss 
the various challenges of final focusing for TeV-scale plasma accelerators and 
propose possible solutions. Finally, we present the first 
design of a final focusing system for a 10 TeV 
linear wakefield collider, evaluate its performance, and 
discuss its shortcomings as well as improvements for 
future designs.

\end{abstract}


\maketitle 

\section{Introduction}


The role of the final focusing system (FFS) in a collider is to 
demagnify the beams to the desired spot sizes at the interaction point (IP) 
in order to maximize the luminosity $\mathcal{L}$: 
\begin{equation}
    \mathcal{L} \propto \frac{1}{\sigma_{x}\sigma_{y}}
\end{equation}
where $\sigma_{x}$ and $\sigma_{y}$ are the root-mean-square beam sizes 
in the $x$ and $y$ planes, respectively. For flat beam collisions with $\sigma_x \gg \sigma_y$, this focusing can be achieved 
by arranging pairs of quadrupole magnets into a telescope, which 
provides point-to-point and parallel-to-parallel imaging~\cite{osti_6281389}. 
Complications arise when considering beams with a finite 
energy spread $\delta$, as 
particles with different energies are focused to different longitudinal 
positions near the IP. These chromatic aberrations, if left untreated, result in 
an increase of the beam size at the IP and degrade the luminosity of the 
collider. Furthermore, the strong focusing fields of the quadrupoles closest to the IP, referred to as the final 
doublet (FD), induce the emission of incoherent synchrotron radiation (SR), which further 
increases the energy spread of the beam, known as the 
Oide effect. 
Schemes for handling the chromatic effects have been 
demonstrated at the SLAC Linear Collider (SLC) \cite{osti_7189349} and Final Focus Test 
Beam (FFTB) \cite{burke_1993} using dedicated chromatic correction sections. A more recent compact 
approach~\cite{PhysRevLett.86.3779} relies on local chromaticity correction by utilizing 
sextupoles interleaved with the FD, decreasing the overall length of the FFS while providing 
larger FD bandwidth than the schemes with dedicated 
chromatic correction sections.
First experimental demonstrations of this approach are being carried out at ATF2~\cite{PhysRevLett.112.034802,PhysRevAccelBeams.19.101001,PhysRevAccelBeams.23.071003,PhysRevAccelBeams.24.051001}. In general, the FFS designs for future colliders need further experimental demonstrations in terms of operational reproducibility, and achieved bunch charge and IP parameters~\cite{arxiv2025}.
A dedicated comparison of FFS designs with the traditional and the compact approaches is presented in reference~\cite{PhysRevSTAB.17.101001}.

Plasma wakefield acceleration (PWFA)~\cite{LindstromReview} has emerged as a leading candidate 
for achieving compact, multi-TeV scale colliders due to the high 
acceleration gradients with respect to state-of-the-art metallic superconducting radio-frequency (SRF) 
technology \cite{Blumenfeld:2007ph,osti_1463003}. The drastic length reduction of the plasma linac may lead to a situation where the
beam delivery system (BDS), which includes the FFS, becomes the largest 
contribution to the facility footprint. The scaling of the BDS system for multi-TeV colliders was examined in ref.~\cite{White2022}. In the FFS 
design with the local compact chromatic correction scheme, the length of the FFS 
scales approximately as \cite{PhysRevLett.86.3779}:
\begin{equation}
    \label{eq:scaling:modernFFS}
    L_{\rm{FFS}} \propto E^{7/10}\, ,
\end{equation}
where $E$ is the energy of the particles. Another 
scheme, derived for scaling the LHC interaction region 
to FCC-hh beam energies under the assumption of 
a constant beam stay clear, utilizes a different 
scaling~\cite{PhysRevAccelBeams.20.081005}:
\begin{equation}
    \label{eq:scaling:clic}
    L_{\rm{FFS}} \propto E^{1/3}\,
\end{equation}
A similar procedure was also utilized for scaling the CLIC 3~TeV FFS to the 7~TeV design~\cite{Manosperti:2023sbd,lewis}. The optimal scaling 
for the FFS length with energy is expected to fall 
somewhere between those shown in Eq.~\ref{eq:scaling:modernFFS} and Eq.~\ref{eq:scaling:clic} by allowing some luminosity loss due to synchrotron radiation. 
As a result, higher energy colliders will require a longer (and thus more 
expensive) FFS. Compact, high-energy linear colliders 
must balance luminosity and length considerations for the FFS. 

The desire for high intensity electron beams at future PWFA colliders also poses a 
problem for the achievable luminosity. Particles in colliding beams  
interact strongly with the field produced by the opposing beam, resulting 
in the emission of SR. This process, known as beamstrahlung, causes each beam to 
radiate away its energy and reduce the center-of-mass energy of the 
colliding particles, broadening the luminosity spectrum~\cite{Yokoya}.

In the following sections we discuss each of these challenges in more 
detail and propose possible solutions. Finally, we present 
the first design of a FFS as part of the 10 TeV Wakefield Collider Design Study~\cite{Gessner:2025acq} and 
assess its performance in the context of a future PWFA 
collider design. 
\section{Beam Size Limiting Processes}
In this section, we review the physical processes that limit our ability to focus beams to arbitrarily small spot sizes. These processes include limitations of beam transport systems as well as dynamic radiative effects.

\subsection{Chromatic Effects}
\label{sec:chromatic}

\begin{figure}[htbp!]
    \centering
    \includegraphics[width=0.50\textwidth]{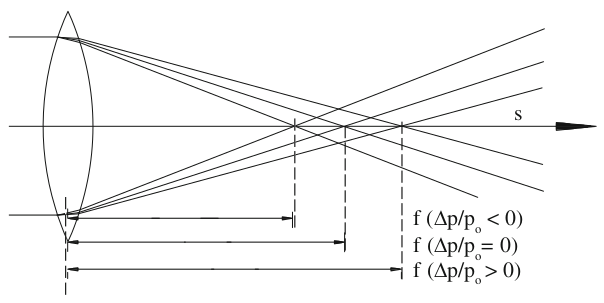}
    \caption{Diagram illustrating the focal points for particles with momentum deviations with respect to the nominal reference energy $p_{0}$. Figure adapted from Ref.~\cite{Wiedemann}.}
    \label{fig:chromatic}
\end{figure} 

In practice, all particle beams have some finite energy spread, 
$\delta$. The bending of an individual particle in a 
magnetic field depends on the energy of 
the particle itself, and, due to this, particles of different energies are 
focused differently in a quadrupole field. This is illustrated in Figure~\ref{fig:chromatic} 
for particles with an energy offset $\Delta p$ with respect to the nominal 
particle energy $p_{0}$; the particles are focused to different 
points along the longitudinal coordinate, which increases 
the spot size at the focal point for the reference 
particle energy (the IP).

Typically, the largest focusing gradients in the FFS are present 
in the final doublet (FD), which demagnify the beams at the IP 
to maximize the luminosity. As a result, the largest contribution 
to the spot size growth due to chromaticity comes from the FD. 
To lowest order, the spot size at the IP for a beam with finite relative energy deviation $
\delta$ can be written as, e.g.~\cite{PhysRevAccelBeams.26.091003}: 
\begin{equation}
    \label{eq:spotsize_dispersion}
    \sigma_{i}^{*} = \sqrt{\epsilon_{i}\beta_{i}^{*}(1+ \xi_i^2\delta^2) + D_{i}^2\delta^2 }
\end{equation}
%
%
where $\epsilon_{i}$ is 
the geometric emittance for the coordinate $i$,  $\beta^*_{i}$ is the beta function for 
coordinate $i$, $D_i$ is the dispersion, and $\xi_i$ is the chromaticity. 
At the IP, design $D_i$ is typically zero and an expression for $\xi_i$ is given in Section~\ref{ssec:10tev:performance}.

Chromatic effects generated by the FD can be compensated by a 
combination of bending magnets to generate dispersion and sextupoles 
to provide transverse-position-dependent kicks. Two main approaches 
for chromatic correction have emerged for linear colliders, differing 
in the arrangement of the bending sections and sextupoles. The 
first scheme developed~\cite{osti_7189349,burke_1993} utilized dedicated sections for 
cancelling the vertical and horizontal chromaticity, each comprised 
of bending magnets with pairs of sextupoles; the sections themselves 
were placed in regions of high dispersion and large-$\beta$. To cancel the 
geometric aberrations generated by the sextupoles, the sextupole pairs are 
arranged such that they are separated by a minus identity transfer matrix. 
A different scheme~\cite{PhysRevLett.86.3779} relies on local chromaticity 
correction by placing sextupoles just before 
and between the FD quadrupoles with a long upstream bending section
to generate the necessary dispersion. 
Horizontal second-order dispersion is 
introduced by the sextupoles, and is corrected by designing the 
optics such that half of the total horizontal chromaticity is generated 
upstream of the FD. Another pair of sextupoles, placed before the bending 
section, are utilized to cancel the geometric aberrations, provided that 
they have the opposite phase of the FD sextupoles. 


\subsection{Oide Effect}
\label{sec:oide}

The strong focusing fields of the FD quadrupoles acting on the beam cause the 
emission of hard SR near the IP, resulting in emittance growth of the beam and 
therefore increasing the beam spot size~\cite{PhysRevLett.61.1713}. With SR emission, the spot 
size at the IP for a linear collider is given by:
\begin{equation}
    \label{sr:spotsize}
        \sigma_{y}^{*2} = 
        \beta_{y}^{*} \epsilon_{y}
        + \frac{110}{3\sqrt{6\pi}} r_{e} \lambda_{e} \gamma^{5} F(\sqrt{K}L, \sqrt{K}\ell^{*}) \left( \frac{\epsilon_y}{\beta_{y}^{*}}\right)^{5/2}
\end{equation}
where $r_{e}$ is the classical electron radius, $\lambda_{e}$ is the 
Compton wavelength, $\gamma$ is the Lorentz factor, and $F$ is a dimensionless function given by:
%
    \begin{align}
        F(\sqrt{K}L, \sqrt{K}\ell^{*}) =
        \int_{0}^{\sqrt{K}L}|\sin\phi + \sqrt{K}\ell^{*}\cos\phi|^{3} \nonumber \\ 
        \times \left[ \int_{0}^{\phi} (\sin\phi' + \sqrt{K}\ell^{*}\cos\phi') \right]^{2} d\phi
    \end{align}
%
where $K$ is the focusing gradient of the quadrupole, $L$ is the thickness of the quadrupole, and $\ell^{*}$ is the distance between the front of the quadrupole and the IP. 
A generalization of the Oide effect considering the two transverse dimensions is presented in~\cite{PhysRevAccelBeams.19.021002} showing a more pronounced growth with the vertical emittance and a dependence of quadratic order on the horizontal emittance.
The function in Equation~\ref{sr:spotsize} has a minimum when $\beta_{y}^{*}$ 
is 
\begin{equation}
    \beta_{y}^{*} = \left[ \frac{275}{3\sqrt{6\pi}} r_{e} \lambda_{e} F(\sqrt{K}L, \sqrt{K}\ell^{*}) \right]^{2/7} \gamma (\epsilon_{Ny})^{3/7}
\end{equation}
This minimum spot size at the IP, referred to as the Oide limit, is given by the 
following expression,
\begin{equation}
    \label{eq:oide_limit}
    \sigma_{y}^{*} = \left( \frac{7}{5} \right)^{1/2} \left[ \frac{275}{3\sqrt{6\pi}} r_{e} \lambda_{e} F(\sqrt{K}L, \sqrt{K}\ell^{*}) \right]^{1/7} (\epsilon_{Ny})^{5/7}
\end{equation}

While the Oide limit has not been directly measured at collider 
experiments, it is a fundamental ingredient in the design of
future proposed colliders, e.g.~\cite{PhysRevAccelBeams.21.011002}. Possible solutions to reduce 
the impact of the Oide effect involve increasing the length or 
decreasing the gradient of the final quadrupole~\cite{Blanco:1967497}, 
and utilizing a pair 
of octupole magnets before and after the final quadrupole~\cite{Blanco-García:2141800}.  
It has also been shown that the 
use of adiabatic 
plasma lenses~\cite{PhysRevLett.64.1231} may mitigate the Oide effect. Further 
discussion of plasma lenses is presented in Section~\ref{sec:plasma_lenses}.

\subsection{Beamstrahlung}

Near the IP, each beam interacts with the electromagnetic field produced by the 
other beam, 
causing energy to be radiated away in the form of SR. This increases the energy 
spread of each beam, allowing for collisions at lower-than-nominal energy and 
resulting in a deterioration of the luminosity while also providing additional background in the detector. The field strength of the beam, 
which characterizes the level of beamstrahlung, is represented by the dimensionless 
parameter $\Upsilon$. For Gaussian beams:  
\begin{equation}
    \Upsilon = \frac{5\gamma N_{b} r_{e}^{2}}{6\sigma_{z}(\sigma_x + \sigma_y)\alpha}
\end{equation}
where $N_{b}$ is the bunch population, $r_{e}$ is the classical electron radius, 
$\sigma_{z}$ is the RMS bunch length, and $\alpha \approx 1/137$ is the fine 
structure constant.

For a fixed beam size at the IP, the problem of beamstrahlung can be greatly 
reduced by the use of flat beams, whereby $\sigma_{x} >> \sigma_{y}$. The 
plasma-based collider designs with collision energy $\sqrt{s} >$ 1 TeV 
operate in the so-called quantum beamstrahlung regime ($\Upsilon >> 1$) due to 
the high collision energy and short bunch lengths. In this regime, 
for fixed luminosity, beam energy, and beam sizes at the IP, the 
number of photons emitted per lepton $n_{\gamma}$ and energy spread 
resulting from beamstrahlung $\delta_{\gamma}$ scale as~\cite{Barklow_2023}:
\begin{equation}
    \label{eq:n_gamma}
    n_{\gamma} \propto \sigma_{z}^{1/3} \Upsilon^{2/3}
\end{equation}
and
\begin{equation}
    \label{eq:delta_gamma}
    \delta_{\gamma} \propto \sigma_{z}^{1/3} \Upsilon^{2/3}.
\end{equation}
Equations
\ref{eq:n_gamma} and~\ref{eq:delta_gamma} suggest that shorter bunch lengths help to reduce deleterious beamstrahlung effects. New studies have examined beamstrahlung as an inevitable aspect of collisions at 10 TeV and explore the physics reach of machines with broad luminosity spectra~\cite{Barklow_2023, HeBeamstrahlung, ChigusaPhysics}.








\section{Studies at 10 TeV}

A nominal flat-beam FFS design for parton center-of-mass (pCM) energies of 10 
TeV is presented here as part of the 10 TeV Wakefield Collider Design Initiative \cite{Gessner:2025acq}. This design is extrapolated from the 
7 TeV FFS for the Compact Linear Collider (CLIC) \cite{lewis}, which employs 
the local chromaticity correction scheme.

\subsection{Scaling CLIC Lattice to 10 TeV}


The increased beam energy of 10 TeV with respect to the CLIC FFS design 
energy of 7 TeV necessitates an energy-dependent scaling to account 
for the change in beam rigidity and the impact of SR on the luminosity.
A procedure, outlined in Ref.~\cite{PhysRevAccelBeams.20.081005}, is performed on the CLIC 
7 TeV FFS lattice and summarized as follows: the length scales and 
normalized magnet strengths are scaled according to a scaling factor 
$\kappa$ related to the ratio of the nominal and new beam energies:
\begin{equation}
    \label{eq:kappa}
    \kappa = \left( \frac{10 \hspace{1mm} \rm{TeV}}{7 \hspace{1mm} \rm{TeV}} \right)^{1/3} \approx 1.126
\end{equation}
The scaling factor is applied for each power of the length in the 
relevant quantities, as shown in Table~\ref{tab:scaling}. This scaling is taken as a 
starting point for a FFS design at 10 TeV; 
future optimizations of the FFS to 
achieve the desired spot sizes at the IP and 
maximize the luminosity at 10 TeV will likely require 
further length scalings of the FFS elements beyond 
what is described above. 

\begin{figure*}[htbp]
    \centering
    \includegraphics[width=1.00\textwidth]{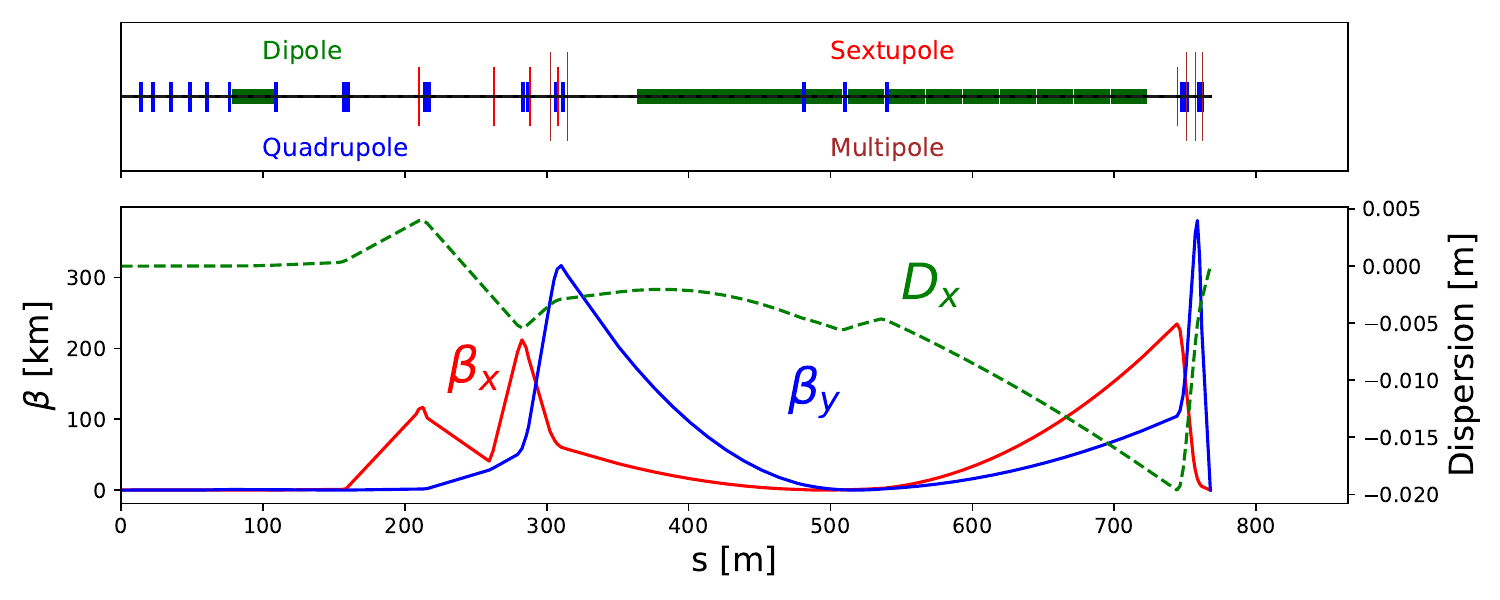}
    \includegraphics[width=1.00\textwidth]{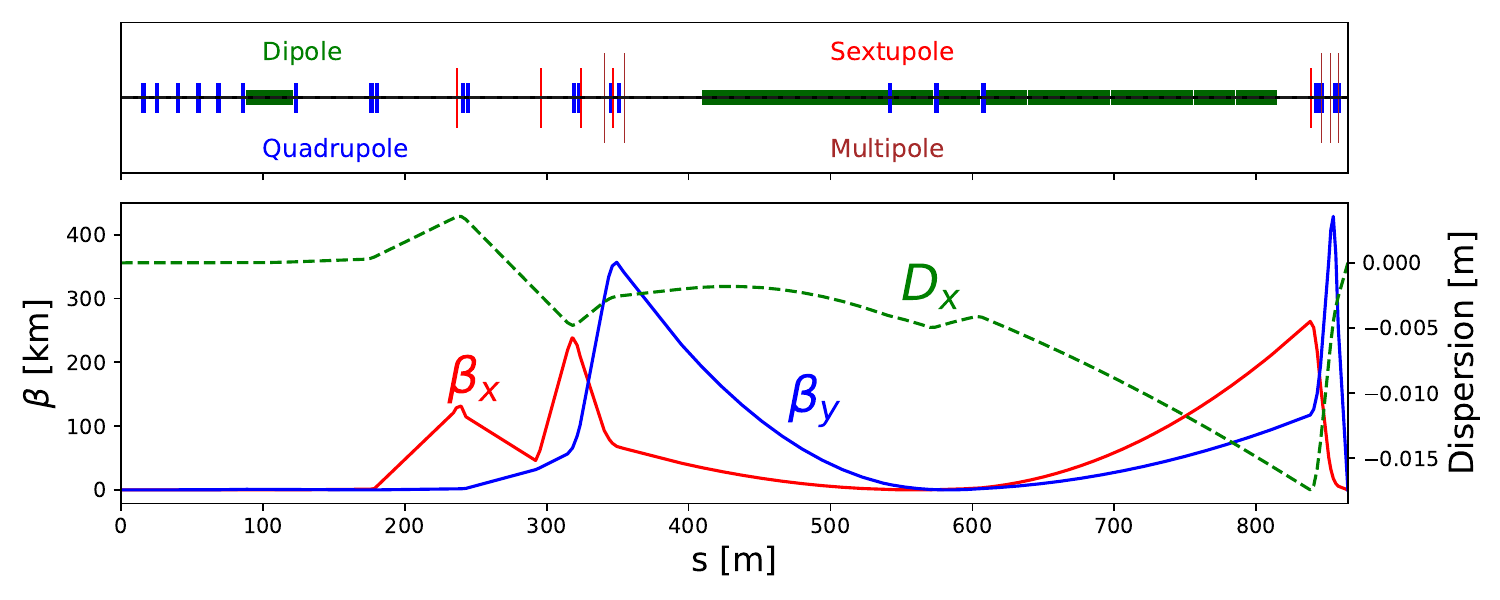}
    \caption{Horizontal and vertical $\beta$ functions and dispersion for 
    the 7 TeV (top) and 10 TeV (bottom) FFS. Surveys of the FFS designs are shown above each plot using the same horizontal scale. The 7 TeV design is 767 meters long and the 10 TeV design is 865 meters long. Dipoles shown in green, 
    quadrupoles in blue, sextupoles in red, and higher-order 
    multipoles shown in brown.}
    \label{fig:beta_functions}
\end{figure*}

\begin{table}[h]
    \centering
      \scalebox{0.9}{
      \begin{tabular}{@{\extracolsep{4pt}}lccccccc@{}}
  \hline\hline
\multicolumn{4}{c}{Scalings of relevant parameters} \\
      \hline
      Quantity & Units & Pre-scaling & Post-scaling \\
      \hline
      Beta Function & [m] & $\beta$ & $\kappa\beta$ \\
      Magnet and Drift Lengths & [m] & $\ell$ & $\kappa\ell$ \\
      Norm. Dipole strength & [$\rm{m}^{-1}$] & $k_{0}$ & $\kappa^{-1}k_{0}$ \\
      Norm. Quadrupole strength & [$\rm{m}^{-2}$] & $k_{1}$ & $\kappa^{-2}k_{1}$ \\
      Norm. Sextupole strength & [$\rm{m}^{-3}$] & $k_{2}$ & $\kappa^{-3}k_{2}$ \\
\hline\hline
  \end{tabular}}
    \caption{Scaling of parameters for increased FFS design energy. The value of the parameter $\kappa$ is given by Eq.~\ref{eq:kappa}.}
    \label{tab:scaling}
\end{table}
Under this scheme, the overall length of the FFS increases from 
767.671 m at 7 TeV to 864.588 m at 10 TeV.
This procedure ensures a constant beam stay clear while maintaining the 
same normalized emittances, resulting in the shortest possible 
interaction region length. The parameters for the 7 TeV and 10 TeV 
FFS designs are summarized in Table~\ref{tab:FFS_params}.
\begin{table}[]
    \centering
      \scalebox{1.0}{
      \begin{tabular}{@{\extracolsep{4pt}}lccccccc@{}}
  \hline\hline
\multicolumn{3}{c}{7 TeV and 10 TeV FFS Parameters} \\
      \hline
      Parameter & 7 TeV & 10 TeV\\
      \hline
      Length [m] & 767.671 & 864.588 \\
      $L^{*}$ [m] & 6 & 6.757 \\
      $\beta_{x}^{*} / \beta_{y}^{*}$ [mm] & 7.00/0.16 & 7.88/0.18 \\
      $\epsilon_{x} / \epsilon_{y}$ [nm] & 660/20 & 660/20 \\
      $\sigma_{x}^{*} / \sigma_{y}^{*}$ [nm] & 34.44/1.55 & 32.40/2.31 \\
      $\sigma_{x,SR}^{*} / \sigma_{y,SR}^{*}$ [nm] & 40.67/3.02 & 43.02/4.21 \\
      $\delta_{p, rms}$ [\%] & 0.3 & 0.3 \\
\hline\hline
  \end{tabular}}
    \caption{Parameters for the 7 TeV and 10 TeV FFS 
    designs.}
    \label{tab:FFS_params}
\end{table}
Figure~\ref{fig:beta_functions} shows the $\beta$ functions 
and dispersions for the 7 TeV and 10 TeV FFS designs; the scaling procedure results in a mild increase in the beta functions at the 
IP. However, for fixed normalized emittance $\epsilon_{i,\rm{N}}$, the increased beam energy results in a decrease in the geometric 
emittance, commonly referred to as adiabatic damping:
\begin{equation}
    \label{eq:emittance}
    \epsilon_{i} = \frac{\epsilon_{i,\rm{N}}}{\beta\gamma}
\end{equation}
where here $\beta=\frac{v}{c}$ is the longitudinal velocity normalized to the speed of light. 
Due to the reduction in geometric emittance, to first order (in the absence of 
dispersion), the beam size after scaling 
$\tilde{\sigma}_{i}$ is approximately reduced by a factor of $\kappa$:
\begin{equation}
    \label{eq:spotsize}
    \tilde{\sigma}_{i} = \sqrt{\frac{\epsilon_{i,\rm{N}}\tilde{\beta}_{i}}{\tilde{\beta}\tilde{\gamma}}} \approx \sqrt{\frac{\epsilon_{i,\rm{N}}(\kappa\beta_{i})}{(\kappa^{3}\beta\gamma)}} = \frac{1}{\kappa}\sigma_{i}
\end{equation}
Consequently, the geometric luminosity should then increase approximately by a 
factor of $\kappa^{2}$. 

\subsection{Lattice Performance}
\label{ssec:10tev:performance}

In reality, the beam size at the IP is comprised of various terms at different 
orders in the transfer map expansion. The transfer map $X$ relates the transverse 
particle coordinates $z_{f}$ at a point along a beamline to the initial coordinates ($x_{i}$,$p_{x,i}$,$y_{i}$,$p_{y,i}$,$\delta_{i}$)
\cite{PhysRevSTAB.9.081001}:
\begin{equation}
    z_{f} = \sum_{jklmn} X_{z,jklmn} x^{j}_{i}p^{k}_{x,i}y^{l}_{i}p^{m}_{y,i}\delta^{n}_{i}\,,
\end{equation}
where $X_{z,jklmn}$ are the map coefficients for the particular final coordinate. 
The order of the map, $M$, is the largest sum of the powers among the considered terms:
\begin{equation}
    M = j + k + l+ m + n\,.
\end{equation}

\begin{figure*}[htbp]
    \centering
    \includegraphics[width=0.80\textwidth]{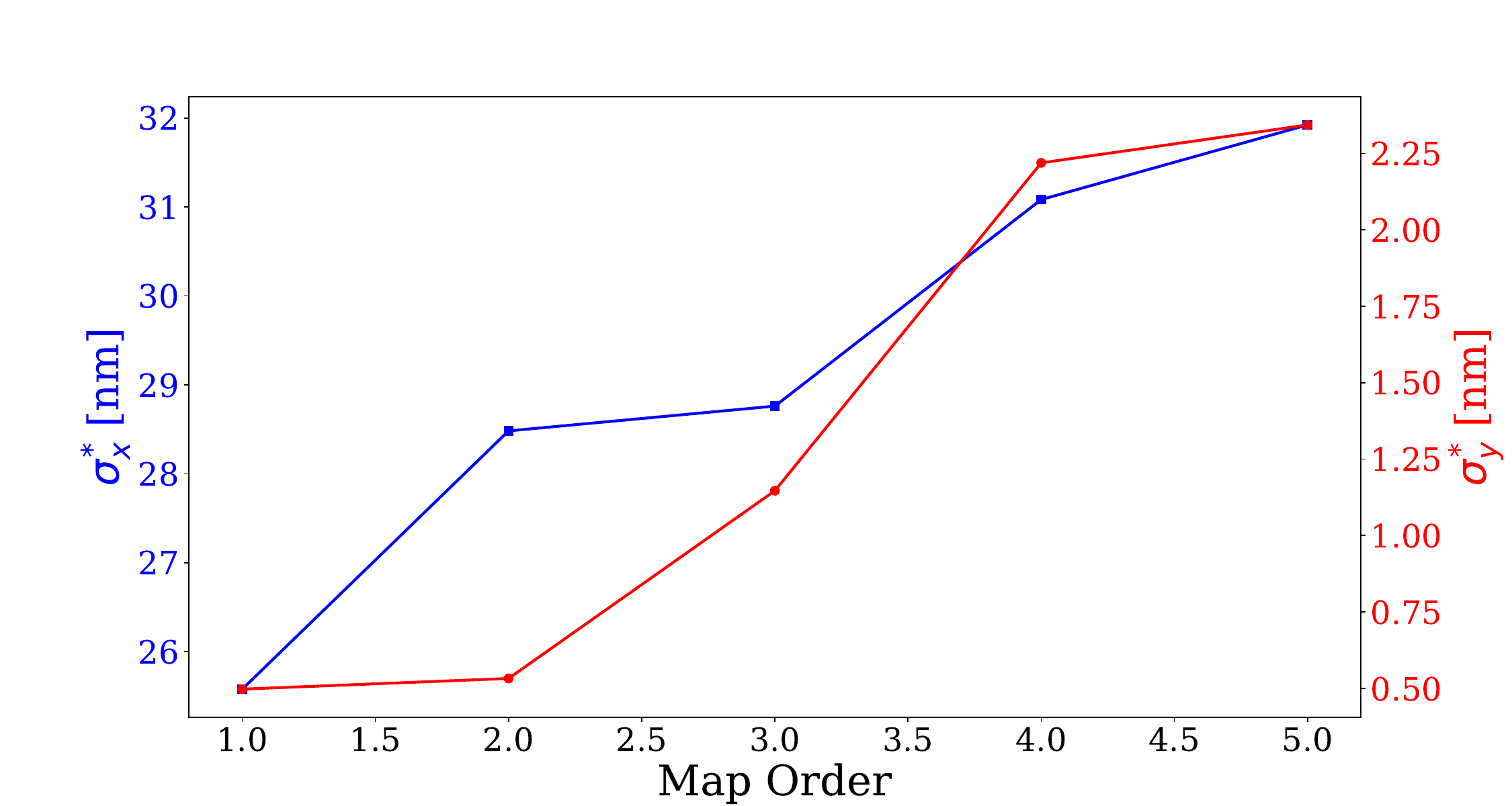}
    \caption{Horizontal (blue) and vertical (red) beam sizes at 10 TeV 
    calculated to different orders in the map. Values calculated 
    using MAPCLASS2.}
    \label{fig:spotsize:maporder}
\end{figure*}

The IP vertical chromaticity $\xi_y$ can be expressed in terms of the map coefficients as~\cite{PhysRevAccelBeams.26.091003}
\begin{equation}
    \xi_y^2 = \frac{1}{\beta^*_y}\left(X_{y,00101}^2\beta_{y0}  + \frac{X_{y,00011}^2}{\beta_{y0}}\right)\,,
\end{equation}
where $\beta^*_y$ is the vertical IP beta function and $\beta_{y0}$ is the beta function at the start of the FFS.
Above $M = 2$, calculation of map coefficients is analytically challenging, 
however many modern particle tracking codes (such as MAD-X~\cite{Grote:618496} with the Polymorphic Tracking Code~\cite{Schmidt:573082}) can numerically calculate the 
coefficients for arbitrary $M$. The spot size at a given $M$ is a complicated 
function of the map coefficients, and can be calculated using the MAPCLASS2 
\cite{Martinez:1491228} program interfaced with MAD-X~\cite{Grote:618496} or the Polymorphic Tracking Code (PTC)~\cite{Schmidt:573082}. For the 10 TeV FFS design, the spot size calculated  as a function of the 
map order is shown in Figure~\ref{fig:spotsize:maporder} up to $M = 5$. The values are compared with 
the spot sizes obtained with the particle tracking code PLACET~\cite{Placet} and 
exhibit good agreement. To evaluate the luminosity with various 
beam-beam effects included, the particle coordinates from tracking 
are fed into GuineaPIG~\cite{GuineaPIG}. The 
luminosity is calculated with and without 
SR enabled in PLACET, and the results are 
shown in Table~\ref{tab:beamsize_lumi}. The impact of 
SR is rather considerable, increasing 
the beam size in the vertical direction by a factor of $\approx 2$ and in 
the horizontal direction by a factor of $\approx 1.25$. 

\begin{figure}[htbp]
    \centering
    \includegraphics[width=0.50\textwidth]{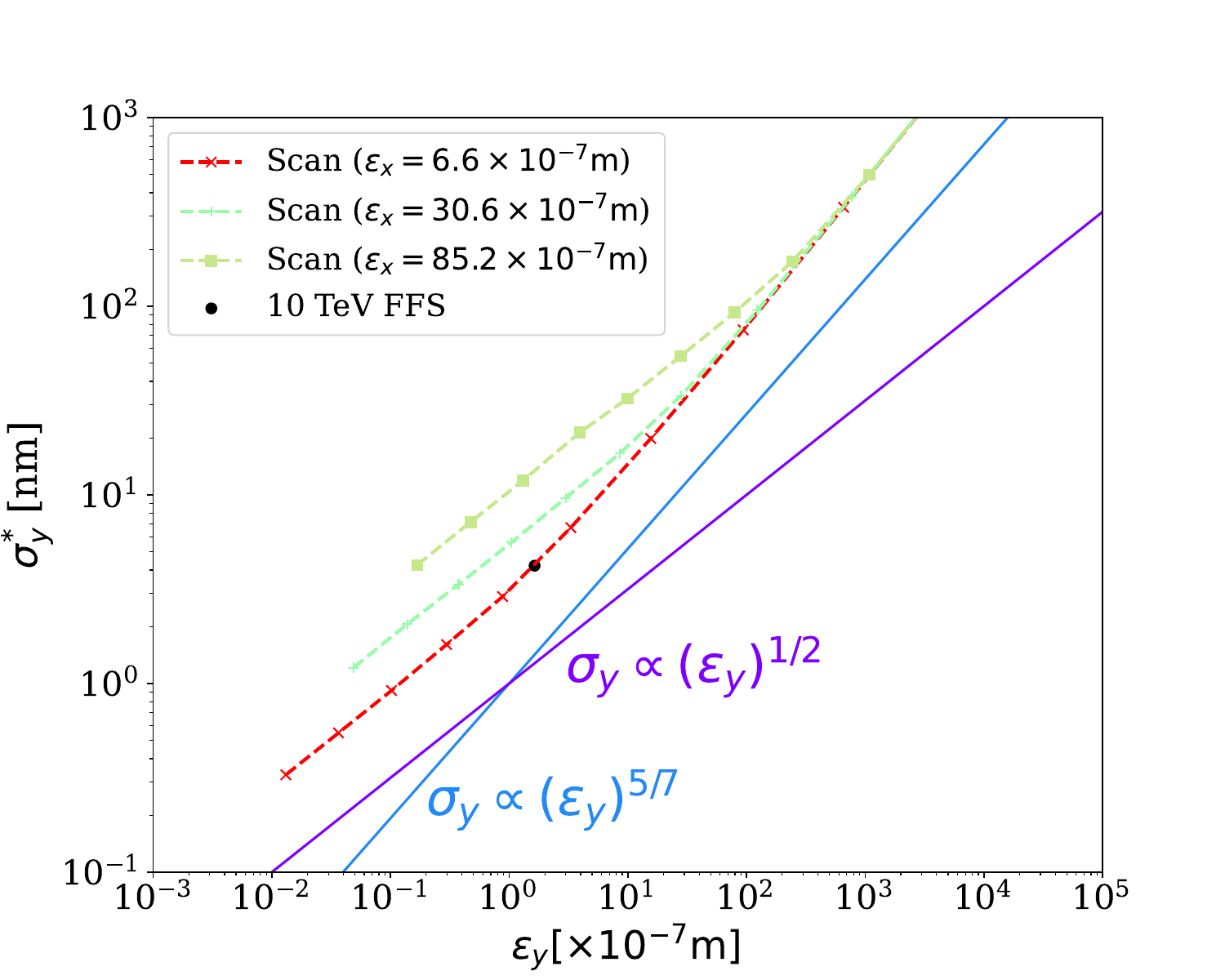}
    \caption{Vertical beam size at the IP as a function of the normalized vertical emittance. Green and red curves show beam size for different values of the normalized horizontal emittance, while the blue and purple lines shown the expected beam size scaling with (Eq.~\ref{eq:oide_limit}) and without (Eq.~\ref{eq:spotsize})
    the Oide effect, respectively. Beam sizes are calculated using PLACET.}
    \label{fig:emittance_scan}
\end{figure}

\begin{figure}[htbp]
    \centering
    \includegraphics[width=0.50\textwidth]{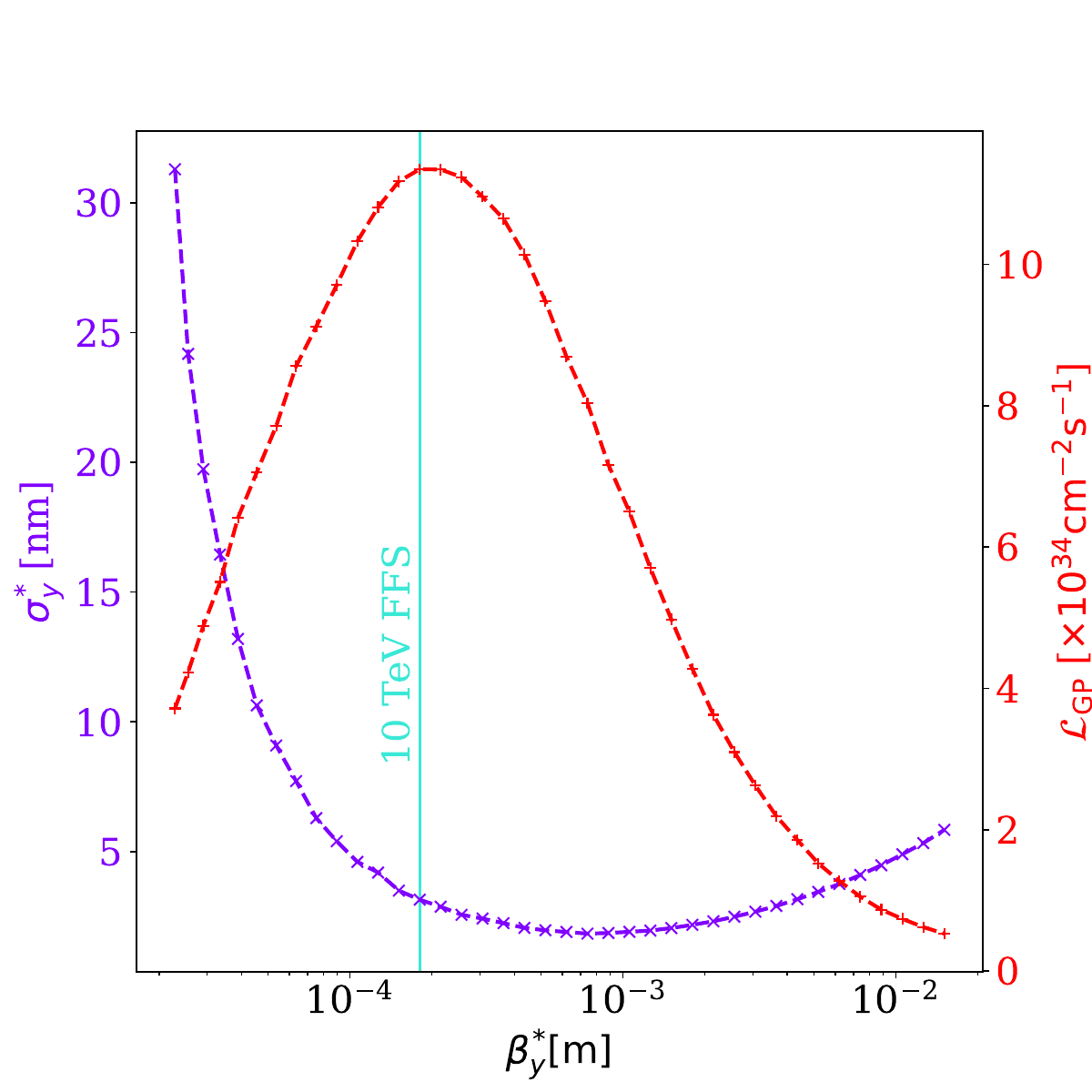}
    \caption{Vertical beam size (purple) and luminosity (red) at the IP as a function of the vertical beta function at the IP. The vertical line shows the nominal beam size for the 10 TeV FFS design.
    Values for $\beta_{y}^{*}$ obtained from MAD-X, while spot sizes and luminosity obtained via PLACET  simulations and GuineaPIG, respectively. }
    \label{fig:beta_scan}
\end{figure}

The sensitivity of the chosen parameters to the 
Oide limit was considered as well. At fixed $\beta_{y}^{*}$, the vertical emittance was scanned 
for different values of the horizontal 
emittance, as shown in 
Figure~\ref{fig:emittance_scan}. The values 
for the 10 TeV FFS, denoted by the black dot, 
is in a region where the Oide effect starts to 
become significant, as the slope of the red 
curve at this point falls between the spot size scaling 
predicted by Equation~\ref{eq:spotsize} 
(purple line) and Equation~\ref{eq:oide_limit} 
(blue line). This behavior is also apparent when varying 
$\beta_{y}^{*}$ for fixed emittances. 
Figure~\ref{fig:beta_scan} shows the evolution of the vertical 
beam size at the IP (purple curve) and luminosity (red curve) for 
different $\beta_{y}^{*}$ values. As indicated by the teal line, the 
$\beta_{y}^{*}$ chosen for the 10 TeV FFS maximizes the luminosity, 
while the minimum $\sigma_{y}^{*}$ is achieved for a 
$\beta_{y}^{*}$ that is $\approx$ 4 times larger. This increase in 
the beam size at peak luminosity results from the fact that the 
beam size is sensitive to tails in the beam that are generated by 
higher-order aberrations and SR from quadrupoles (e.g. Oide effect), 
while the luminosity is highly dependent on the core of the beam 
distribution.


\begin{table}[]
    \centering
      \scalebox{1}{
      \begin{tabular}{@{\extracolsep{4pt}}lccccccc@{}}
  \hline\hline
      & $\sigma_{x}^{*}$ [nm] & $\sigma_{y}^{*}$ [nm] & $\mathcal{L}_{\rm{GP}}$ [$ 10^{34} \rm{cm}^{-2} \rm{s}^{-1}$] \\
      \hline
      Without SR & 32.40 & 2.31 & 19.2 \\
      With SR & 43.02 & 4.21 & 10.8 \\
      \hline
      Target Values & 18.4 & 0.45 & 34.1 \\
\hline\hline
  \end{tabular}}
    \caption{Beam size and Luminosity for the 10 TeV FFS with and without synchrotron radiation enabled and comparison with idealized 
    target values adapted from Ref.~\cite{Barklow_2023}. Spot sizes 
    calculated using PLACET and Luminosity calculated using GuineaPIG.}
    \label{tab:beamsize_lumi}
\end{table}

\section{Takeaways and Future Studies}

\begin{figure}[htbp]
    \centering
    \includegraphics[width=0.50\textwidth]{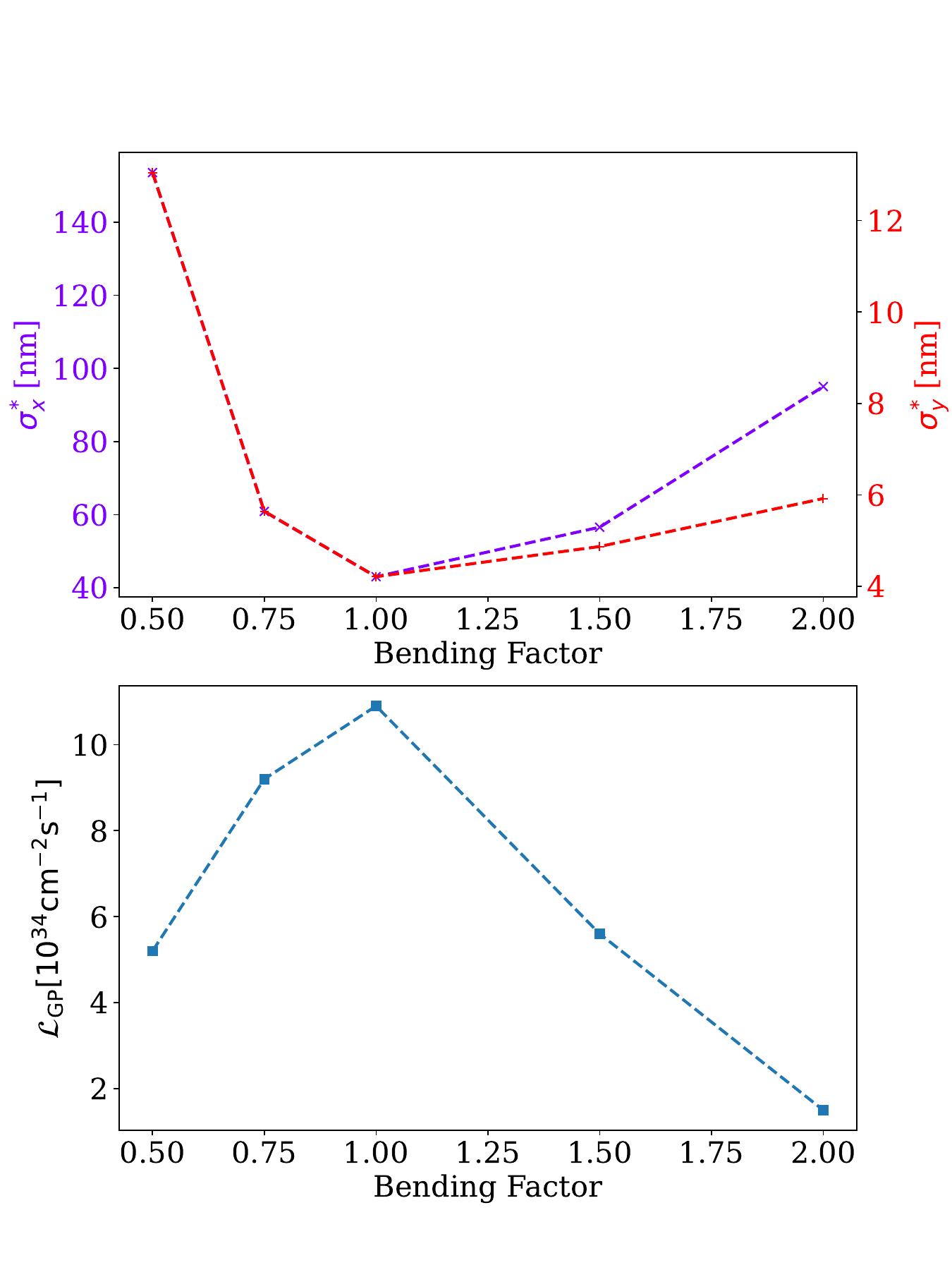}
    \caption{Beam sizes (top) and luminosity (bottom) for different values of the bending factor.
    Values for the spot sizes and luminosity were obtained via PLACET simulations and GuineaPIG, respectively. }
    \label{fig:bending_scan}
\end{figure}

The 10 TeV FFS lattice presented here is the first 
design of a FFS design for a linear collider operating 
at a parton center of mass energy of 10 TeV. However, 
there are various improvements that can be made to 
maximize its performance. The increase in the 
vertical and horizontal beam sizes 
due to SR can be reduced by increasing the lengths and 
decreasing the gradients of the final focusing 
quadrupoles. Additionally, the contribution to the 
horizontal spot size growth from SR in the bending 
sections may be reduced by increasing the lengths or 
decreasing the bending angles of the dipoles, with 
corresponding adjustments to the sextupole strengths to 
maintain small horizontal dispersion at the IP. To investigate this, a 
bending factor was introduced as a multiplier to the dipole strength 
and the inverse of the sextupole strength. Figure 
\ref{fig:bending_scan} shows the spot sizes (top) and luminosity (bottom) as a 
function of this bending factor, with a maximum luminosity and 
minimum IP spot size occurring at the nominal 
value. This suggests possible higher-order aberrations contributing 
to the decrease in luminosity, requiring careful study of the 
higher-order optics to mitigate these effects. 

In the context of the 10 TeV wakefield collider design 
study~\cite{Gessner:2025acq}, this first 10 TeV FFS design falls short of the luminosity target of $\mathcal{L} = 34.1\times10^{34} \hspace{1mm} \rm{cm}^{-2} \hspace{1mm} \rm{s}^{-1}$ for a flat-beam PWFA collider outlined by the 
Snowmass 2021 report~\cite{Barklow_2023}. Adjusted to 
10 TeV from the initial 15 TeV proposed design energy 
for a flat-beam PWFA collider, the target beam sizes 
in the vertical and horizontal directions are 
0.45 nm and 18.4 nm, respectively; a factor of 
$\approx 10$ smaller in y and $\approx 2.5$ smaller in x. 
It should be noted that these reported target values were computed without any consideration for realistic beam dynamics and thus represent an idealized scenario. Even with the 
proposed improvements stated above, to achieve this 
decrease in the beam size, a more dedicated redesign of 
the lattice is likely required. 


Planned future studies include the design of a round-beam FFS 
at 10 TeV, leveraging the fact that PWFA is easier to achieve 
with round beams. For this, two designs will be considered: one 
consisting of conventional focusing elements while the other 
incorporating plasma lenses into the FFS to attempt to 
reduce its overall length and improve focusing capabilities. 
A discussion on the properties of plasma lenses is presented in 
Section~\ref{sec:plasma_lenses}. 

Finally, we should note that the traditional Beam Delivery System (BDS) in conventional linear collider designs includes several additional subsystems beyond the Final Focus and diagnostic sections. These include the beam switch yard (for dual interaction points or tune-up beam dump beamlines) and, most notably, the energy and betatron collimation sections.

The characteristics and length of the collimation sections depend on various design decisions and assumptions, such as the use of survivable versus consumable spoilers, the expected fractional beam population in the transverse and energy tails, and many other factors. As seen in earlier BDS designs---such as that of CLIC---at energies above 1~TeV, the collimation system can become the dominant component in terms of length. This occurs because, under typical design assumptions, its scaling with energy is less favorable than that of the final focus system itself.

Consequently, for a 1-10~TeV center-of-mass plasma-based linear collider to remain competitive, the design assumptions for the collimation system must be fundamentally reconsidered. Future design work on collimation optimized for plasma colliders will include, but not be limited to, concepts such as distributed collimation integrated within the plasma acceleration sections, modified approaches to protecting the experimental detector from backgrounds, and related ideas. The development of an optimized collimation system will be an important component of our future studies.


\subsection{Plasma Lenses}
\label{sec:plasma_lenses}

The thin, underdense, passive plasma lens~\cite{PhysRevAccelBeams.22.111001} 
is a promising alternative 
to conventional quadrupole magnets for focusing particle beams.
Unlike quadrupoles, plasma lenses can provide axisymmetric, 
geometric-aberration-free focusing, allowing 
for pairs of quadrupoles to be replaced by a single plasma lens. In 
addition, plasma lenses can achieve focusing gradients many orders of 
magnitude larger than those achieved by quadrupoles. For a thin plasma lens 
of non-uniform number density $n_{0}(z)$, the focal length is given by the following 
expression~\cite{PhysRevAccelBeams.22.111001}:
\begin{equation}
     f = \frac{1}{KL} =  \frac{1}{2\pi r_{e}} \frac{\gamma}{\int n_{0}(z) dz} 
\end{equation}
A comparison of the focal lengths for a plasma lens of density 
$n_{0} = 10^{17} \rm{cm}^{-3}$ and quadrupole magnets for a 5 TeV 
electron beam is presented in Table~\ref{tab:focusing_strengths}. These 
properties of the plasma lens are highly attractive, as they are useful 
for reducing the overall length of the FFS while providing smaller 
beam spot sizes at the IP.



A particularly useful application of the plasma lens 
in the FFS of a TeV-scale collider would be the replacement of quadrupole 
pairs (such as the FD) in order to shrink the overall length needed for the 
FFS~\cite{PhysRevD.40.923}. Although plasma lenses offer high focusing gradients, the hard SR 
emission (Oide effect) can be avoided by shaping the lens such that it 
acts as an adiabatic focuser~\cite{PhysRevLett.64.1231}. 
This provides a great advantage over conventional quadrupole focusing 
near the IP as it allows for smaller achievable spot sizes. 
Furthermore, 
in the adiabatic regime, the focusing provided by plasma lenses is 
achromatic; particles of different energies are focused to the same 
point. Consequently, this property could alleviate the need for 
chromatic correction in the FFS and reduce the overall length by eliminating the need for long bending sections and sextupoles. 

Despite their numerous advantages over conventional focusing elements, plasma lenses do pose significant challenges to implement in future 
colliders. Precise shaping of the plasma lens to achieve adiabatic 
focusing is difficult to achieve in practice, and this remains an 
active field of research. For colliders with high repetition rates (kHz or above), 
plasma lenses may create concerns as to their robustness, as the plasma recovery time may 
result in density profile variations on a shot-to-shot basis, leading to different 
focusing properties. It has also been demonstrated that the driving of plasmas using flat beams can lead to asymmetric focusing fields experienced by the witness bunch and variation 
in the focusing fields along the longitudinal 
coordinate~\cite{28c4-blhg}, 
necessitating precise experimental control and matching to 
prevent chromatic beam size growth and proper focusing along both 
axes. 
Additionally, the desire for adiabatic 
focusing would increase the plasma density closer to the IP, resulting 
in beam-plasma scattering that could result in emittance growth~\cite{PhysRevLett.64.1231} and background events in the detector~\cite{PhysRevD.40.923}. Finally, despite successful experimental demonstrations of plasma lensing of positron beams~\cite{Ng2001}, it remains an open question as to whether or not positrons can be focused to small spot sizes using a plasma lens without geometric aberrations~\cite{Muggli2008}.

\begin{table*}[hbtp!]
    \centering
      \scalebox{1}{
      \begin{tabular}{@{\extracolsep{4pt}}lccccccc@{}}
  \hline\hline
      Focusing Element & $K$ $[\rm{m}^{-2}]$ & $L$ $[\rm{mm}]$ & $f$ $[\rm{cm}]$ \\
      \hline \hline
      Superconducting Quadrupole: G = 100 T/m & 0.06 & 410 & 4100 \\
      \hline
      Permanent Magnetic Quadrupole: G = 500 T/m & 0.3 & 180 & 1800 \\
      \hline
      Underdense Plasma Lens: $n_{0} = 10^{17} \rm{cm}^{-3}$ & 180 & 7.4 & 74 \\
\hline\hline
  \end{tabular}}
    \caption{Comparison of different focusing elements for a 5~TeV electron beam assuming a uniform phase advance $\Delta\psi = \sqrt{K}L = 0.1$ across the element.}
    \label{tab:focusing_strengths}
\end{table*}




\section{Conclusions}
\label{sec:conclusions}

We present the first design of a 10 TeV final focusing system by scaling up the 
version utilized for CLIC 7 TeV, resulting in a luminosity of 
$1.08\times 10^{35}$ $\rm{cm}^{-2}$ $\rm{s}^{-1}$. The reduction to the 
luminosity due to synchrotron radiation is found to be considerable, and we also
demonstrate that our chosen $\beta$-functions and emittances are sensitive to the 
Oide limit. Future mitigation studies to reduce the impact of synchrotron 
radiation and the Oide effect are discussed, requiring a more careful 
redesign of the final focusing system. Other plans for the beam delivery system are considered, including the implementation of a round-beam final focusing system utilizing passive plasma lenses, the addition of a 
collimation section, and the further optimization of higher-order beam aberrations to maximize the luminosity. 

\bibliographystyle{unsrtnat}
\bibliography{references}











\newpage

\end{document}